\documentstyle[aps,multicol,epsf,epsfig]{revtex}

\begin{document}

\draft
\tightenlines

\title{Reply to ``Comment on `Long-range electrostatic interactions between like-charged
colloids: steric and confinement effects' ''}

\author{Emmanuel Trizac$^1$ and Jean-Luc Raimbault$^2$}

\address{
$^1$ Laboratoire de Physique Th\'eorique\cite{umr1}, B\^atiment 210, Universit\'e 
de Paris-Sud, 91405 Orsay Cedex, France\\
E-mail: Emmanuel.Trizac@th.u-psud.fr\\
$^2$ Laboratoire de Physique et Technologie des Plasmas, \'Ecole Polytechnique,
Route de Saclay, 91128 Palaiseau \\
E-mail: raimbault@lptp.polytechnique.fr
}
 
%\date{\today}

\maketitle
\begin{abstract}
In his Comment (cond-mat/0104060) to [Phys. Rev. E 60, 6530 (1999)], Mateescu shows that
while the effective interactions remain repulsive when the specific size
of the micro-ions is taken into account via a Modified Poisson-Boltzmann
equation, a similar conclusion cannot be reached for the situation of 
complete lateral confinement. This point is correct but has already been 
considered in a more
general study [Phys. Rev. E 62, R1465 (2000), where repulsion 
is generically obtained]; moreover, we argue that it illustrates the 
irrelevancy of the notion of pair potential in completely confined 
configurations, as shown on a simple example.
\end{abstract}

\begin{figure}
\vspace{-1cm}$$\input{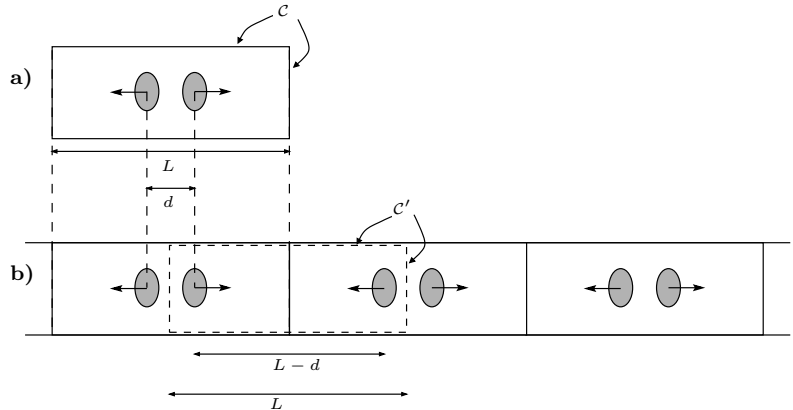}$$
%\vskip 1cm
\vspace{-5mm}\caption{The two electrostatically equivalent configurations 
a) two colloids immersed in a cylinder ${\cal C}$ of length $L$
with mirror symmetry;
b) the series of aligned image colloids (every mid-plane between two adjacent
colloids is a plane of symmetry for the whole charge distribution). 
The cylinder ${\cal C}'$ of length
$L$ is indicated by the thick dashed line. In ${\cal C}$, the distance between the 
colloids is $d$ whereas in ${\cal C}'$, this distance is $L-d$. 
The straight simple arrows indicate the direction of the 
effective force acting on the
colloids (repulsive in ${\cal C}$ and consequently attractive in
${\cal C}'$).
}
\label{fig}
\end{figure}

\begin{multicols}{2}
\vskip 5mm
In his Comment \cite{Mateescu} to our work \cite{Trizac}, E.M. Mateescu 
establishes that the situation of finite lateral confinement does not necessarily 
lead to a repulsive effective pair potential between like charged colloids in an
electrolyte, thus invalidating one of the conclusions reached in \cite{Trizac}.
Although correct, this precise point has already been 
considered and incorporated in the more
general analysis presented in \cite{Pre} where we focussed on the limit 
of infinite lateral extension for the confining cylinder, and obtained rigorously 
a generic
effective repulsion \cite{Rqe}. Moreover, it is worthwhile to stress that 
the counter-example provided by Mateescu illustrates the inapplicability of the
pair potential concept, as becomes clear below. Consider the situation of two colloids
confined in a finite-length cylinder ${\cal C}$ 
with Neumann boundary conditions 
(vanishing normal electric field). The method of image charges allows to construct 
the equivalent infinite series of image colloids as depicted in Fig \ref{fig}.
If we assume the effective interactions repulsive in ${\cal C}$ (as in Fig \ref{fig}), 
it is sufficient to notice that the boundary conditions on ${\cal C}'$ are also
of Neumann type to obtain that the effective forces in the cylinder ${\cal C}'$ are 
attractive. Conversely, attractive interactions in ${\cal C}$ correspond
to effective repulsions in ${\cal C}'$. This shows that with Neumann boundary
conditions, any repulsive configuration can be mapped onto an attractive
one in a closely related cell, and that the resulting effective attraction/repulsion
is a spurious effect of image charges. Consequently, the situation of 
complete lateral confinement is irrelevant when discussing the sign of an 
effective pair potential.

\end{multicols}


\begin{thebibliography}{99}
\bibitem{Rqe} In this limit, Eq. (8) of Ref. \cite{Pre} is equivalent 
to relation (7) in \cite{Mateescu}.
\end{thebibliography}
\end{document}